\documentclass[aps,preprint,amsmath,amssymb,nofootinbib]{revtex4-1}
\usepackage{xcolor}
\usepackage{graphicx} 
\usepackage[caption=false]{subfig}

\usepackage{gensymb} 
\usepackage{amsmath} 
 \usepackage[english]{babel}
 \makeatletter\AtBeginDocument{\let\@elt\relax}\makeatother
\usepackage{braket} 
\usepackage{booktabs} 
\usepackage{ulem}
\usepackage[colorlinks=true]{hyperref} 
\hypersetup{colorlinks,
	citecolor=blue
}
\usepackage[capitalize]{cleveref}

\newcommand{\ba}{\begin{eqnarray}}
\newcommand{\ea}{\end{eqnarray}}

\begin{document}


\title{Superscaling in the resonance region for neutrino-nucleus scattering: The SuSAv2-DCC model.}


\author{J. Gonzalez-Rosa$^a$, G. D. Megias$^a$, J. A. Caballero$^{a,b}$, M. B. Barbaro$^{c,d}$}
\affiliation{$^a$Departamento de F\'isica At\'omica, Molecular y Nuclear, Universidad de Sevilla, 41080 Sevilla, Spain}
\affiliation{$^b$ Instituto de F\'{\i}sica Te\'orica y Computacional Carlos I, Granada 18071, Spain}

\affiliation{$^c$Dipartimento di Fisica, Universit\`a di Torino, Via P. Giuria 1, 10125 Torino, Italy}
\affiliation{$^d$INFN, Sezione di Torino, Via P. Giuria 1, 10125 Torino, Italy}



\date{\today}

\begin{abstract}
In this work the SuSAv2 and dynamical coupled-channels (DCC) models have been combined and tested in the inelastic regime for electron and neutrino reactions on nuclei. The DCC model, an approach to study baryon resonances through electron and neutrino induced meson production reactions, has been implemented for the first time in the SuSAv2-inelastic model to analyze the resonance region. Within this framework, we also present a novel description about other inelasticities in the resonance region (SoftDIS). The outcomes of these approaches are firstly benchmarked against ($e,e^{\prime}$) data on $^{12}$C. The description is thus extended 
 to the study of neutrino-nucleus inclusive cross sections on $^{12}$C and $^{40}$Ar and compared with data from the T2K, MicroBooNE, ArgoNEUT and MINERvA experiments, thus covering a wide kinematical range.  
		
\end{abstract}


\maketitle

\section{Introduction}\label{Introduction}

The description of neutrino-nucleus reactions is essential for the analysis of neutrino oscillation experiments and the determination of relevant properties such as the violation of the charge-parity symmetry in the neutrino sector 
and 
the mass hierarchy~\cite{alvarez-ruso_nustec_2018}. These studies are of paramount relevance to reduce one of the leading experimental systematics, the cross section and neutrino flux determination, which is strongly related to nuclear-medium uncertainties. Most of past, current and future experiments  - MiniBooNE, MicroBooNE, T2K, NOvA, MINERvA, ArgoNEUT, DUNE and HyperK~\cite{abratenko_first_2019-1,acero_measurement_2023,minera_collaboration_double-differential_2020,ruterbories_measurement_2021,t2k_collaboration_measurement_2018, abe_measurement_2020-1, acciarri_measurements_2014-2, miniboone_collaboration_first_2010,dune_collaboration_long-baseline_2016, argoneut_collaboration_first_2012, proto-collaboration_hyper-kamiokande_2018} - operate in the 0.5-10 GeV region, where different channels play a relevant role in the nuclear response. 
The quasi-elastic (QE) regime, associated to one-nucleon knockout, is a very prominent contribution in the range from hundreds of MeV to a few GeV  of initial neutrino energy.
In this region it is also necessary to consider the emission of two nucleons, 
denoted as 2p2h (two-particle-two-hole) channel, and the resonance (RES) regime, corresponding to the excitation of nucleonic resonances followed by their decay and the subsequent emission of pions and other mesons. As the neutrino energies increase up to several GeV, not only the resonance regime but also other inelasticities, corresponding to non-resonant meson production and deep-inelastic scattering (DIS) processes, become more relevant. 
This energy domain 
is of interest for some of the above-mentioned experiments such as MINERvA or ArgoNEUT, and will be essential for the next-generation DUNE experiment. Although most of current measurements are focused on the CC$0\pi$ (or "quasielastic-like") channel, which is defined as charged-current (CC) reactions with no pions ($0\pi$) detected in the final state and, thus, dominated by QE and 2p2h contributions, the inelastic region is also 
accessed 
in some of these experiments via CC-inclusive measurements, where 
no specific hadronic final state is selected, and hence
all reaction mechanisms 
have to be considered. The inelastic regime, which includes resonant and non-resonant meson production and deep-inelastic scattering, can also represent an important background in CC$0\pi$ data.

The resonance 
regime has been extensively studied in previous works by different groups~\cite{PhysRevC.69.035502,PhysRevC.71.015501,Sato_2009, PhysRevC.68.032201}. Emphasis has been placed not only on the description of the nucleonic resonances but also on the treatment of the nuclear effects introduced in the analysis of lepton-nucleus reactions.
Moreover, there is a lack of accurate models and specific measurements 
in the so-called Shallow Inelastic Scattering (SIS) region, that is, the transition region between the resonant and DIS regimes. Information about the resonant nucleon form factors and the inelastic structure functions is mainly extracted from electron scattering data, which implies some restrictions when extended to the neutrino case, as the axial channel is missing in electron reactions. This extension thus requires relying on different approximations based on QCD calculations, quark models, and Parton Distribution Functions (PDFs) or semi-phenomenological models. Nonetheless, most of these approaches are affected by kinematical restrictions and large uncertainties, which makes it difficult to get a consistent and accurate description of the full inelastic regime. 
The SIS and DIS channels 
have been investigated in several recent studies~\cite{gonzalez-jimenez_pion_2018-1,gonzalez-jimenez_nuclear_2019-2,martini_unified_2009-1,vagnoni_inelastic_2017-2, ivanov_superscaling_2012-1,singh_nuclear_1998-1, praet_ensuremathdelta-mediated_2009-1, barbaro_susa_2021-1, amaro_using_2005-2, barbaro_inelastic_2004-2,ivanov_charged-current_2016-1} 
but, for the moment, no satisfactory models have been fully developed in the kinematics of interest for oscillation experiments.

More specifically, several groups \cite{REIN198179,FOGLI1979116,lalakulich_resonance_2006,schreiner__1973} have 
studied pion production in nuclei, providing different descriptions of 
the initial nuclear state, pion production in a bound nucleon, and the possible subsequent pion-nucleon interaction within the residual nucleus. Most of the initial studies were based on the simple Fermi gas approach of non-interacting nucleons, but recently more sophisticated descriptions have been developed, incorporating relativistic mean field nuclear potentials, Random Phase Approximation calculations or spectral functions. 
Regarding resonant production in the nucleon, several groups have also developed sophisticated approaches to analyze the nucleon structure in this regime, such as the MK model~\cite{Kabirnezhad_2018,Kabirnezhad:2022znc} or the Dynamical Coupled-Channels (DCC) model~\cite{nakamura_dynamical_2015-3,Nakamura:2018ntd, DCConline}, which have been tested against electron and neutrino scattering data.

In a recent work~\cite{gonzalez-rosa_susav2_2022}, the superscaling model SuSAv2, initially developed for CCQE neutrino-nucleus cross sections, was extended to the full inelastic regime (SuSAv2-inelastic), where the resonance production and deep inelastic contributions were 
described via the extension to the neutrino sector of the SuSAv2 inelastic model developed for ($e, e'$) reactions. 
The model merges inelastic structure functions~\cite{the_nnpdf_collaboration_unbiased_2011,bjorken_asymptotic_1969,tooran_qcd_2019-1, bodek_experimental_1979-1, bodek_fermi-motion_1981-1,callan_high-energy_1969,ccfr_collaboration_notitle_1997,kim_notitle_1998,ghermann_notitle_1999}, coming from QCD analyses, parton distribution functions (PDFs) or phenomenological approaches, with a mean-field nuclear scaling function to describe 
the nuclear dynamics. This approach also allows one to discriminate between different inelastic regions by introducing some restrictions in the allowed final-state invariant mass. 
For example,  a semi-phenomenological $\Delta$ resonance model based on the scaling function extracted from ($e, e'$) data was also developed and combined with the SuSAv2-inelastic model after removing this resonant contribution 
from the latter. The comparison with both electron and neutrino data was rather satisfactory.

With the aim of studying the sensitivity of these results to different inputs for the elementary lepton-nucleon inelastic structure functions, in this work we implement the DCC model for pion production in lepton-nucleon interactions within the SuSAv2-inelastic nuclear framework above described. The DCC model has been widely tested for electron and neutrino scattering off a single nucleon, 
and has  the merit of considering a rather complete description of the resonant and non-resonant regimes, also including the interaction between the different resonance channels ($\pi N$, $\pi\pi N$, $\eta N$, $K \Lambda$, $K \Sigma$), the interference between resonant and non resonant amplitudes and the neutrino induced two-pion production. 
In the new approach presented in this paper we thus implement the accurate description of pion production in the resonant and non-resonant regimes provided by the DCC model, while using the inelastic structure functions employed in the former SuSAv2-inelastic model at higher kinematics (SIS and DIS).  The combination of the SuSAv2 framework and the DCC model is defined in this work as SuSAv2-DCC or RES-DCC. In the latter, we explore different prescriptions for these inelastic structure functions. This framework also allows us to quantify the contribution of the DIS/SIS channel 
to the resonant and non-resonant regimes 
by subtracting the RES-DCC  contribution from the full inelastic prediction. This will be defined as "Soft-DIS" in this manuscript 
and is connected with the SIS region. 

All these approaches are described in Sect.~\ref{Theoretical}, where the theoretical formalism for the inelastic regime is also summarized  and a comparison of the DCC parametrization and the different inelastic structure functions available in the SuSAv2-inelastic framework is  presented.
 In Sect.~\ref{Results} we show the comparison of our 
 predictions with data:  in Sect.~\ref{Electron-scattering} the analysis of electron reactions  on $^{12}$C is shown as a first benchmark to test the validity of the approaches before applying them to the neutrino case; in Sect.~\ref{Neutrino-scattering} a comparison with CC-inclusive neutrino cross-section measurements on $^{12}$C and $^{40}$Ar is presented at different kinematics and for several experiments. In Sect.~\ref{Conclusions} we draw our conclusions.

\section{Theoretical Background}\label{Theoretical}

The superscaling approach (SuSA) is based on the scaling properties exhibited by inclusive electron scattering where the QE scattering cross section can be written, under certain conditions, as a term containing the single-nucleon cross section times a scaling function ($f$) that embodies the nuclear dynamics. 
The analysis of inclusive electron scattering data~\cite{donnelly_superscaling_1999-2} has shown that, for not too low transferred momentum ($q$ larger than about 400 MeV/c), the scaling function does not depend on $q$ (scaling of 1st kind) nor on the nuclear species (scaling of 2nd kind) and can therefore be expressed in terms of a single variable $\psi$, the so-called scaling variable.
 A more detailed description of superscaling can be found in Refs.~\cite{donnelly_superscaling_1999-2,amaro_electron-_2020-1, amaro_neutrino-nucleus_2021-1, megias_inclusive_2016-1, megias_charged-current_2016-1, megias_neutrinooxygen_2018-1, megias_analysis_2019-2, megias_phd_2015}. 
 This approach has been also successfully applied to inclusive CCQE neutrino scattering and, most recently, to the full inelastic regime for both electron and neutrino reactions. The description of the 2p2h channel has also been included in the model on the basis of the fully relativistic calculation of Ref.~\cite{Simo_2017,AMARO2011151}. 
The corresponding model for the quasielastic region (SuSAv2-QE) is based on a set of QE scaling functions extracted from the relativistic mean field (RMF) model for the nucleus.

 The SuSAv2-inelastic model is an extension of the  SuSAv2-QE approach to the inelastic regime~\cite{gonzalez-rosa_susav2_2022,megias_charged-current_2017}. 
 The double differential inclusive cross section for lepton-nucleus scattering with respect to transferred energy $\omega$ and the scattered lepton solid angle $\Omega$ can be written in the very general form~\cite{amaro_electron-_2020}
 \begin{equation}
\frac{d\sigma}{d\Omega d\omega} = \sigma_0 \sum_K v_K R_K \,,
 \end{equation}
 where $\sigma_0$ is an elementary cross section (the Mott cross section in the case of electron scattering), $v_K$ are kinematic Rosenbluth factors and $R^K$ are the nuclear response functions, containing all the nuclear dynamics. The summed index $K$ is associated to different components of the nuclear tensor with respect to the direction of the transferred momentum ${\bf q}$.
 %
The  nuclear responses depend on the transferred momentum and energy $(q,\omega)$ 
and on the invariant mass $W_{X}\equiv m_N\mu_X$ of the hadronic final states. In the SuSAv2-inelastic model they are given by 


\begin{widetext}
\begin{equation}\label{RKinel}
R^{inel}_{K}(q,\omega,W_X)=N\,\frac{2T_Fm_{N}^{3}}{k^{3}_{F}q}\,\int_{\mu_{X}^{min}}^{\mu_{X}^{max}}d\mu_{X}\mu_{X}f^{model}(\psi_{X})G^{inel}_{K},
\end{equation}
\end{widetext}
being $N$ the number of nucleons participating to the reaction, $k_{F}$ the Fermi momentum and $T_{F}\equiv \sqrt{m_N^2 + k_{F}^{2}} - m_N$ the Fermi kinetic energy.
Thus, the inelastic nuclear responses are defined as the integral  over all  possible final  hadronic states of the single-nucleon inelastic hadronic tensor $G^ {inel}_{K}$ times the inelastic scaling function $f^{model}$ evaluated in a given nuclear model. The latter is written in terms of $\psi_X\equiv\psi_X(q,\omega,W_X)$, which is the extension of the QE scaling variable $\psi$ to the inelastic regime and now depends on the final state invariant mass $W_X$.
The limits of the integral ($\mu_{X}^{min/max}$) depend on the kinematics, on the 
specific inelastic channel (full inelastic, DIS, RES, etc.) and on the range of validity of the inelastic structure functions used to evaluate the single-nucleon tensor. These limits can also be altered 
to mix different models, avoiding double counting. 

\begin{figure*}[!htbp]
  \includegraphics[width=\textwidth]{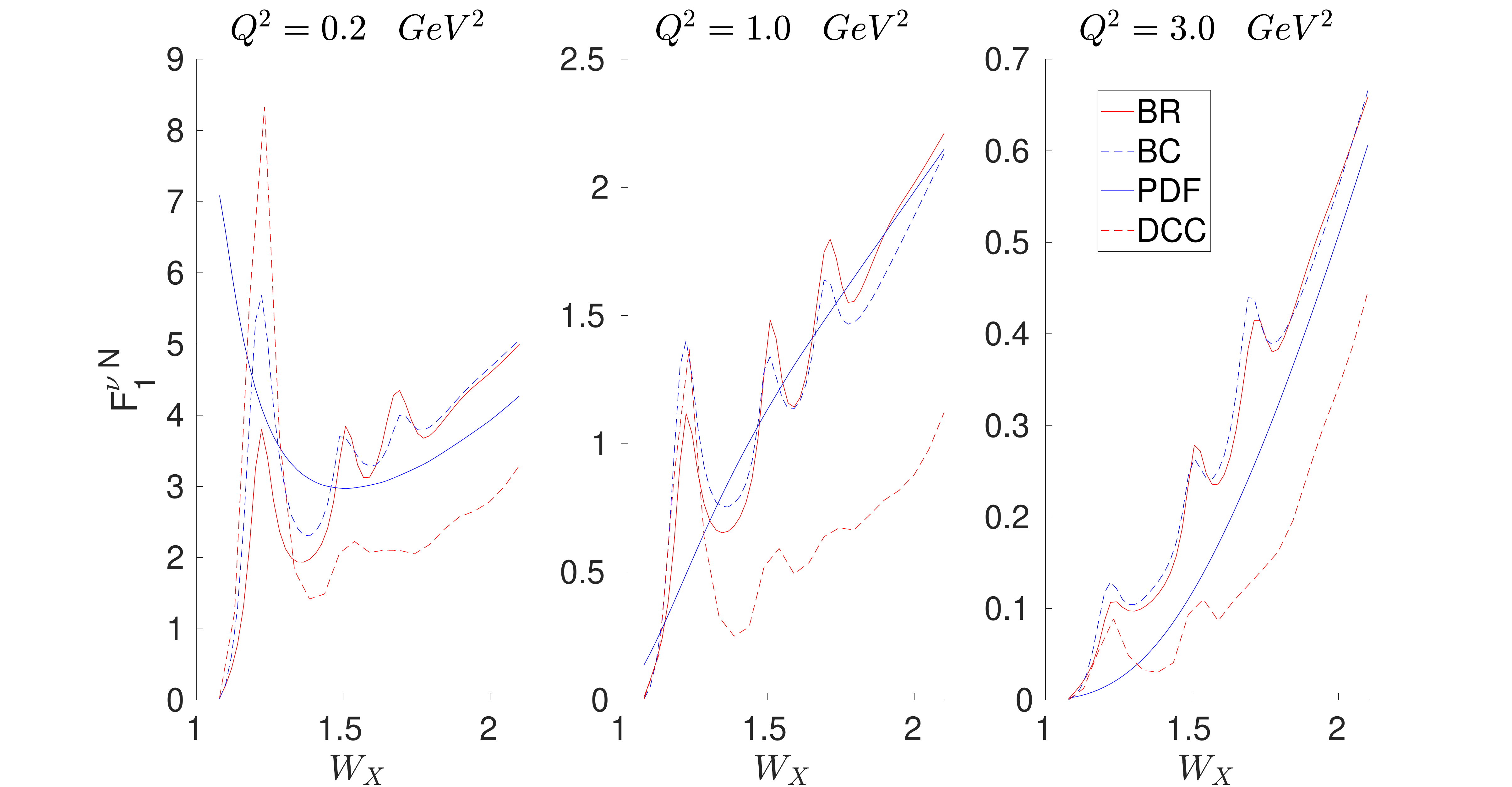}
    \caption{Neutrino inelastic structure function ($F_{1}$) in function of the invariant mass. At various $Q^{2}$-values 0.2 (left), 1.0 (center) ans 3.0 (right) GeV$^{2}$. We show Bodek-Ritchie (BR), Bosted-Christy (BC), parton distribution functions (PDF) and dynamical coupled-channels (DCC) parametrizations  \label{str_function} .}
\end{figure*}

In previous work~\cite{gonzalez-rosa_susav2_2022,megias_inclusive_2016} we have employed either phenomenological inelastic structure functions (Bodek-Ritchie (BR), Bosted-Christy (BC)) or parton distribution functions (GRV98)~\cite{callan_high-energy_1969-1, bodek_axial_2013-1,tooran_qcd_2019-2, bodek_fermi-motion_1981-1, bodek_further_1981-1, bodek_experimental_1979-2, stein_electron_1975-2, liang_notitle_2004, gluck_dynamical_1998-1, bosted_empirical_2008-1} to analyze the full inelastic regime, 
and we have also addressed their extension to the neutrino sector. However, the ranges of validity of these approaches are quite different from each other and not all of them are accurate to describe the resonant and non-resonant regimes. 
Specifically, BC works well for $Q^{2}<8$ GeV$^{2}$ and $1.1<W_{X}/{\rm GeV}<3.1$, PDF for high $Q^{2}$ ($0.8<Q^{2}/GeV^{2}<10^{6}$ )  and $W\gtrsim 3$ GeV, and BR covers a $Q^{2}$ range from 0.1 to 30 GeV$^{2}$. The above-mentioned DCC model for pion production shows a range of validity within $W_X<2.1$ GeV and $Q^{2}<3.0$ GeV$^{2}$. In Fig.~\ref{str_function}, 
where the inelastic neutrino-nucleon structure function $F_1$ is displayed versus the invariant mass, the differences between these approaches can be clearly noticed:  BR and BC describe the resonant structures observed in the inelastic regime and moderate kinematics and a monotonic tail corresponding to DIS, while PDF just predict an average of the resonance region~\cite{Malace_2009}.
On the contrary, the DCC model only describes pion production and the corresponding curves rapidly decrease after  the resonance region as no deep-inelastic scattering contributions are considered, being thus consistently below the predictions of the other approaches.


Accordingly, in this work we split the inelastic contributions in three parts. First, to address resonant and non-resonant pion production, we make use of the DCC model, setting the limits of the integral of Eq.~\ref{RKinel} to  $W_{X}^{min}=m_{N} + m_{\pi}$  and $W_{X}^{max}=2.1$ GeV, according to the validity range given by \cite{nakamura_dynamical_2015-3}. Above $W_{X}=2.1$ GeV, we make use of the inelastic structure functions mentioned in Sec. \ref{Introduction} to describe the DIS regime. 
Furthermore, Fig.~\ref{str_function} clearly shows there
are still inelastic processes not accounted for by the DCC approach below $W_{X}=2.1$ GeV. In this case, we consider DIS contributions both above the resonance region described by DCC (``TrueDIS") and within this resonance region (``SoftDIS"). For TrueDIS, the limits are 
 $W_{X}^{min}=2.1$ GeV
 and 
$W_{X}^{max}=m_N + \omega - E_{s}$, being $E_{s}$ the separation energy. The SoftDIS contribution shares the limits of the resonance region but, in order to avoid double counting, we subtract the contribution from the SuSAv2-DCC model in the Soft-DIS results:
\begin{widetext}
\begin{eqnarray}
\left(\frac{d^{2}\sigma}{d\Omega dk_{l}} \right)_{\rm SoftDIS}=\left(\frac{d^{2}\sigma}{d\Omega dk_{l}} \right)_{inelastic}^{W_{X}^{min}=m_N + m_{\pi}; W_{X}^{max}=2.1\,{\rm GeV}} 
- \left(\frac{d^{2}\sigma}{d\Omega dk_{l}} \right)_{\rm RES-DCC}.
\end{eqnarray}
\end{widetext}

In the following section, we compare the results obtained for electron and neutrino reactions using the SuSAv2-inelastic nuclear model together with the different approaches considered to describe the nucleon dynamics.

\section{Results} \label{Results}

The description of the electron and neutrino inclusive scattering processes requires taking into account the contribution of different reaction channels. These contributions and their abbreviations as stated in the legends  are shown in Table \ref{Table} as well as the model used to describe them. At low values of 
transferred energy ($\omega \approx Q^2/2m_{N}$), the dominant process is quasielastic and it is described by the SuSAv2 superscaling model. As the value of $\omega$ increases, a process can occur in which  2p2h states are excited via meson exchange. This is described by using the relativistic Fermi gas as a framework (RFG-MEC). At higher values of $\omega$, we observe a series of resonances that are modeled by the SuSAv2-DCC discussed in Sec. \ref{Theoretical}. In the same region, it is possible that the neutrino/electron interacts with the partons (SoftDIS), which is taken into account by a combination of SuSAv2-DCC and SuSAv2 inelastic. At even higher energy transfers, the deep inelastic scattering interactions are described by the SuSAv2-inelastic model (TrueDIS).      

\begin{table*}
\begin{tabular}{c c c}
\hline
\hline
Abbreviation & Contribution & Model \\
\hline
 QE    & Quasielastic &  SuSAv2 superscaling \\
  MEC   & 2p2h excitations   & RFG-MEC  \\
  RES  &   Resonances  &   SuSAv2-DCC \\
  SoftDIS &  Deep inelastic scattering ($W_{X}<2.1$ GeV) &  SuSAv2 inelastic; SuSAv2-DCC  \\
  TrueDIS &  Deep inelastic scattering ($W_{X}>2.1$ GeV)  & SuSAv2 inelastic   \\
  All Contr. & All Contributions & \\ 
  \hline
  \hline
\end{tabular}
 \caption{Channels that contribute to the reaction mechanism with the notation followed in the text and the model used to evaluate the cross section.\label{Table}}
 \end{table*}


In Sec \ref{Electron-scattering}, we show our prediction for double-differential inclusive electron-carbon cross section separated in different contributions and 
compared with experimental data. Subsequently, in Sec \ref{Neutrino-scattering} different neutrino-carbon/argon inclusive cross sections are shown and compared with experimental data from T2K, MINERvA, MicroBooNE and ArgoNEUT.

\subsection{Electron scattering } \label{Electron-scattering}

In this section we test the models with inclusive 
$(e,e')$ scattering data. 
\begin{figure*}[!htbp]
  \includegraphics[width=\textwidth]{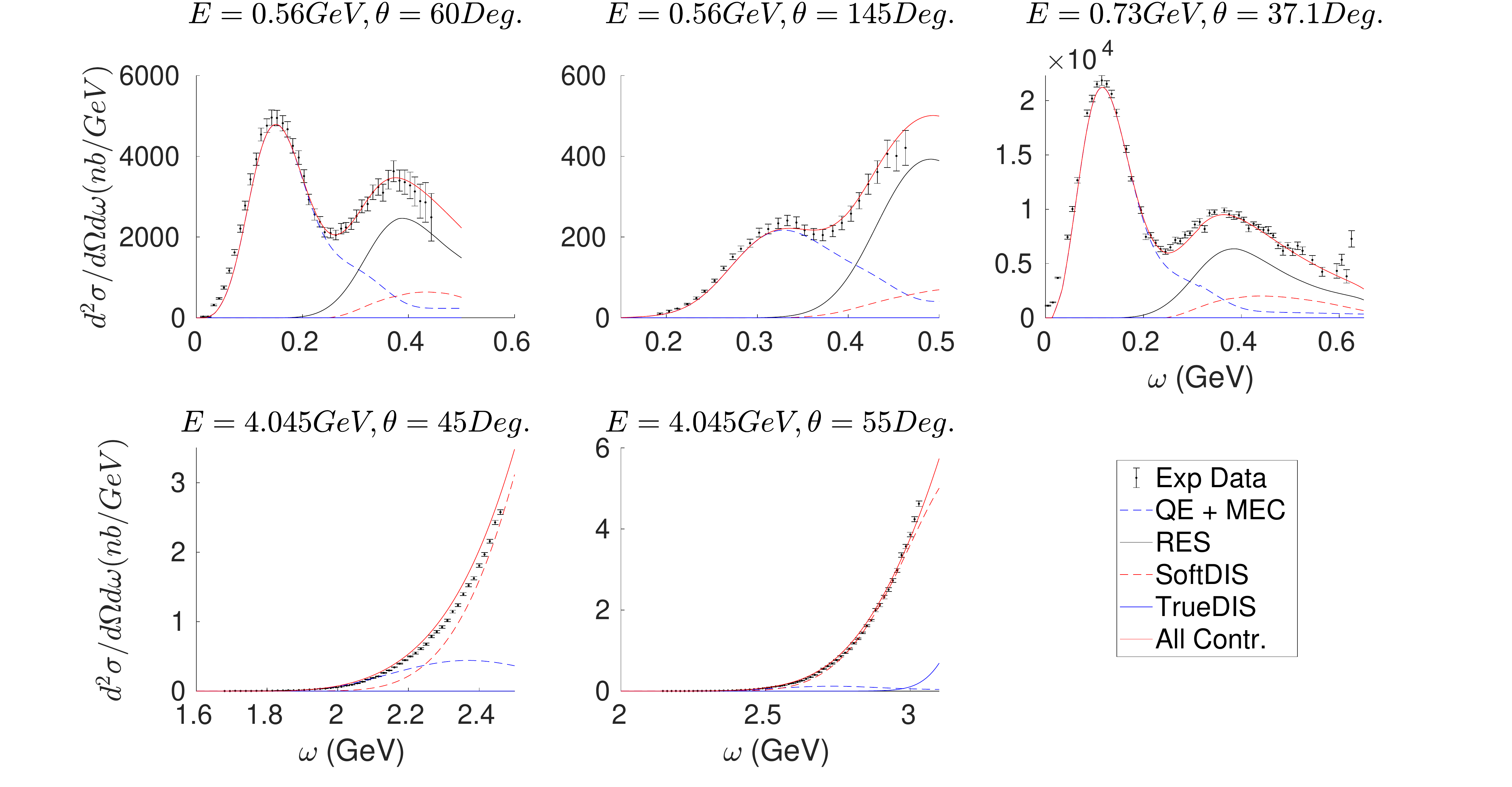}
    \caption{Double-differential inclusive cross section for e-$^{12}$C scattering at given energies and scattering angles (labeled in the panels). It is displayed in function of the transferred energy. Legend explained in  Table \ref{Table}. Data taken from \cite{https://doi.org/10.48550/arxiv.nucl-ex/0603032}.
    \label{Electron} }
\end{figure*}
The double differential electron-carbon cross section versus the energy transfer, $\omega$, is shown in Fig.~\ref{Electron}. In the top panels ($E=560 - 730$  MeV) we observe at lower transferred energy a peak dominated by the quasielastic contribution that is well reproduced by the SUSAv2-QE model. At higher $\omega$, one enters into the dip region where 
two-particle two-hole meson exchange current contributions are needed. The second peak observed corresponds to the excitation of a $\Delta$-resonance. In this region the contribution provided by SuSAv2-DCC and SoftDIS is in good agreement with data. In the two bottom panels ($E=4045$ MeV, $\theta=45,55$ Deg.), the largest contribution corresponds to SoftDIS that is evaluated by including a combination of SuSAv2-inelastic  and SuSAv2-DCC. As shown, TrueDIS is negligible in all cases with the exception of a minor increase observed at $\omega > 3$ GeV in the right-bottom panel. 
This is consistent with the quite low values of the transfer energy considered. Notice that TrueDIS is only relevant at very high $\omega$ values.

Summarizing, we have shown that the different models considered in this work are capable to provide a precise description of electron scattering data, and therefore, we extend their use to neutrino scattering processes.

\subsection{Neutrino scattering} \label{Neutrino-scattering}

We now compare our results with several CC inclusive neutrino scattering data. To assess the accuracy of the predictions for double differential cross sections, for each data set we perform a $\chi^{2}$  analysis making use of the measured covariance matrix. This matrix 
measures the correlations between different bins of data. We obtain the $\chi^{2}$ as follows:
\begin{widetext}
\begin{eqnarray}
 \chi^{2}_{i,j}&=&(x_{i, measured} - x_{i, expected})V_{ij}^{-1}(x_{j, measured} - x_{j, expected})
\label{eq:chi2ij}
\\
 \label{eq:chi2}
\chi^{2}&=&\sum_{i}\sum_{j}\chi^{2}_{i,j}\,,
\end{eqnarray}
\end{widetext}
where the indices $i,j$ denote a given pair of data bins, $x_{measured}$ is the cross section  measured in the experiments, $x_{expected}$ is the cross section predicted by the models and $V$ is the measured covariance matrix provided by the experiments.
In the following analysis for neutrinos we will show,  for each set of data, both the value of $\chi^{2}_{i,j}$ for each data bin \eqref{eq:chi2ij} and the sum of all the $\chi^{2}$ values \eqref{eq:chi2}. 

For the single-differential cross section, the $\chi^{2}$-studies are done using the following formula: 
\begin{equation}
    \chi^{2}=\sum_{i}\left(\frac{x_{i,expected}-x_{i,measured}}{\Delta x_{i,measured}}\right)^{2},
    \label{chi2-sc}
\end{equation}
where 
$\Delta x_{i,measured}$ is the uncertainty associated to the measurements.  In the description of the figures we specify which definition are we using.

\subsubsection{T2K}
\begin{figure*}[!htbp]
  \includegraphics[width=\textwidth]{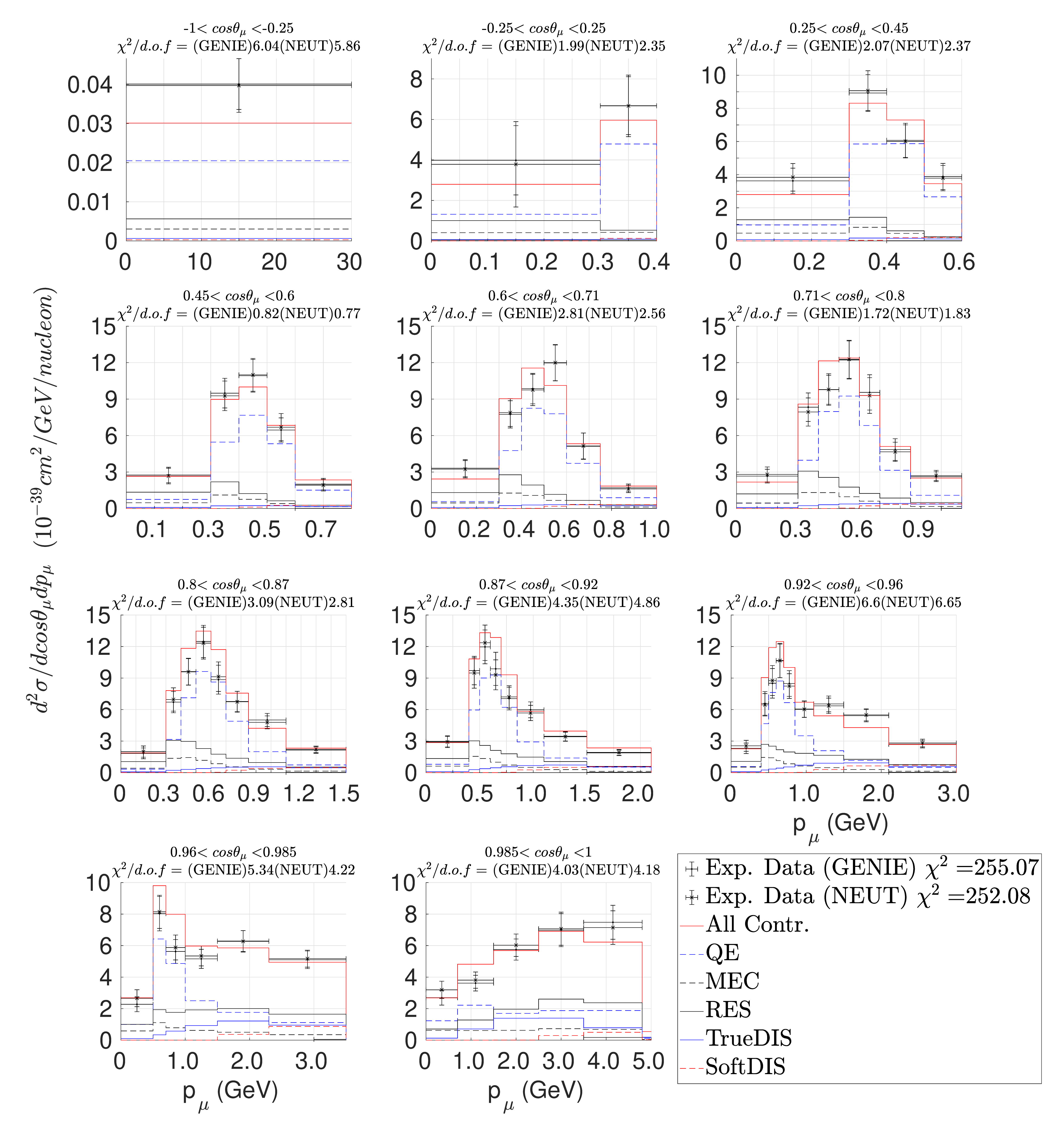}
    \caption{T2K CC inclusive flux-averaged double-differential cross section per target nucleon in bins of the muon scattering angle (labeled in panels) as function of the muon momentum. The different contributions are shown individually. Also, we show the sum of all of them (see Table \ref{Table}). Data taken from \cite{t2k_collaboration_measurement_2018}. The $\chi^{2}$- value shown in each panel is a partial calculation associated to each bin. We are using Eq. \ref{eq:chi2} to calculate the result portrait in legend. \label{T2K} }
\end{figure*}

\begin{figure*}[!htbp]
  \includegraphics[width=\textwidth]{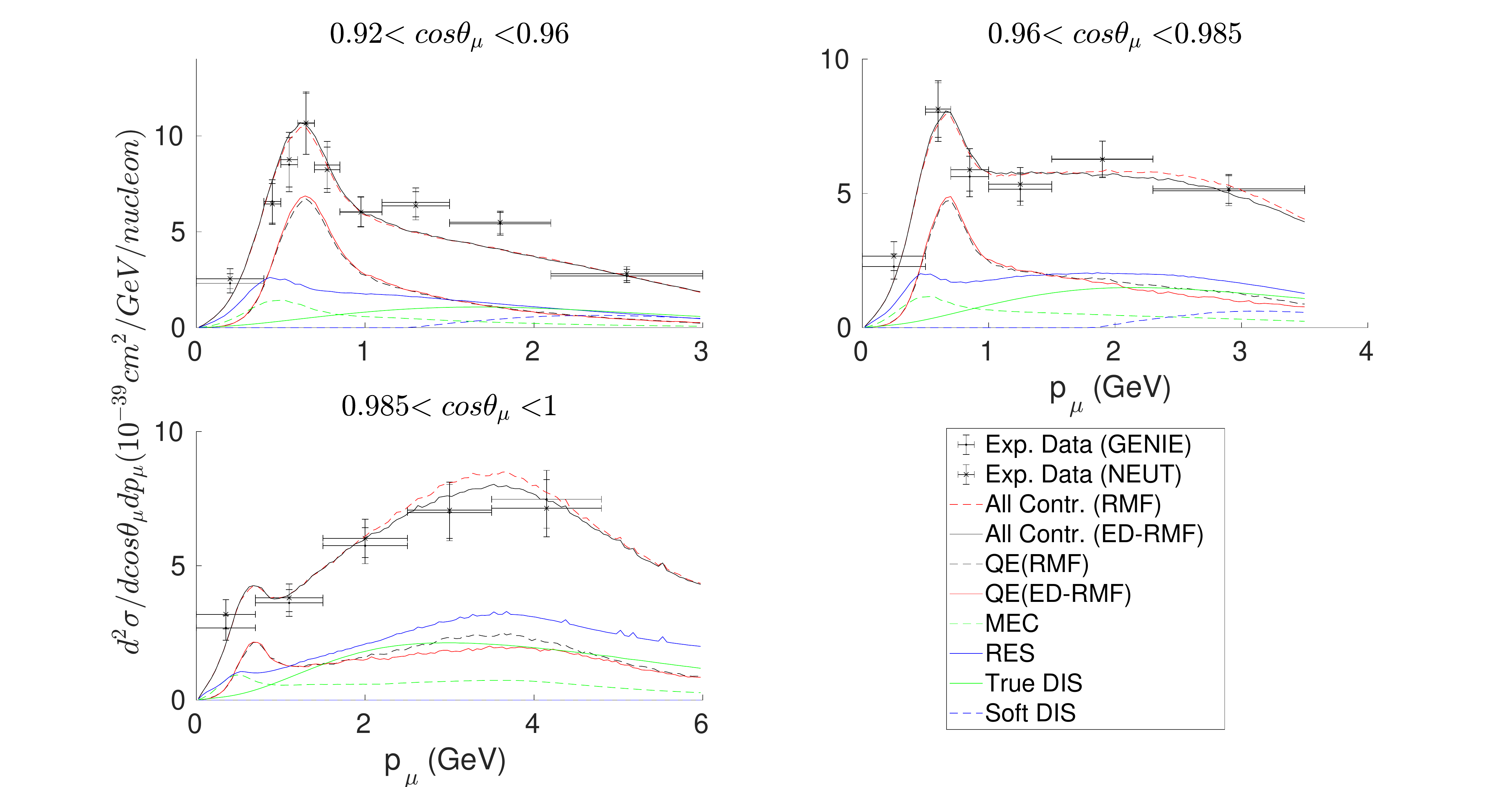}
    \caption{T2K CC inclusive flux-averaged double-differential cross section per target nucleon in bins of the muon scattering angle (labeled in panels) as function of the muon momentum. QE contribution obtained using RMF and ED-RMF as labeled in the legend. Legend as in 
    Fig.~\ref{T2K} (see Table \ref{Table}). Data taken from \cite{t2k_collaboration_measurement_2018}. 
    \label{RMF} }
\end{figure*}

In the T2K experiment, the target used is carbon and the neutrino flux peaks at $\sim$0.6 GeV.  In Fig.~\ref{T2K} the CC inclusive $\nu_{\mu}$ - $^{12}$C double differential cross section per nucleon is displayed for various angular bins as a function of the muon momentum $p_{\mu}$.
In each panel two sets of data are shown, corresponding to experimental analyses performed using the GENIE~\cite{GENIE:2021npt} and NEUT~\cite{Hayato:2009zz,Hayato:2021heg} generators.
The theoretical predictions for the cross section are folded with the T2K flux and the different channels are shown separately. In general, the sum of QE and MEC processes gives  about 40\% of the cross section and provides the biggest contribution followed by RES. The resonance contribution varies between roughly 20 and 
35\% and decreases as the scattering angle gets larger. TrueDIS and SoftDIS are not very important above  $\sim$30$^{\circ}$ of scattering angle. However, at lower angles and for muon momentum above 1.5 GeV these contributions become crucial in order to reproduce the experimental data, being TrueDIS the most important of the two although SoftDIS becomes relevant to account for the higher $p_\mu$ experimental data in the last panel. It can be observed that at  momentum below 1.5 GeV some data points are overestimated. 
This excess can be fixed by using Relativistic Mean Field (RMF) or Energy Dependent Relativistic Mean Field (ED-RMF) models, as it is confirmed in Fig.~\ref{RMF} for the last three angular bins. 
Qualitatively the results
are very similar to the ones found in our previous work~\cite{gonzalez-rosa_susav2_2022}, where the SuSAv2-inelastic model was used. 
Moreover, the $\chi^{2}$-values are very similar to the ones presented in the simulations with GENIE and 
NEUT~\cite{t2k_collaboration_measurement_2018}.


\begin{figure*}[!htbp]
  \includegraphics[width=\textwidth]{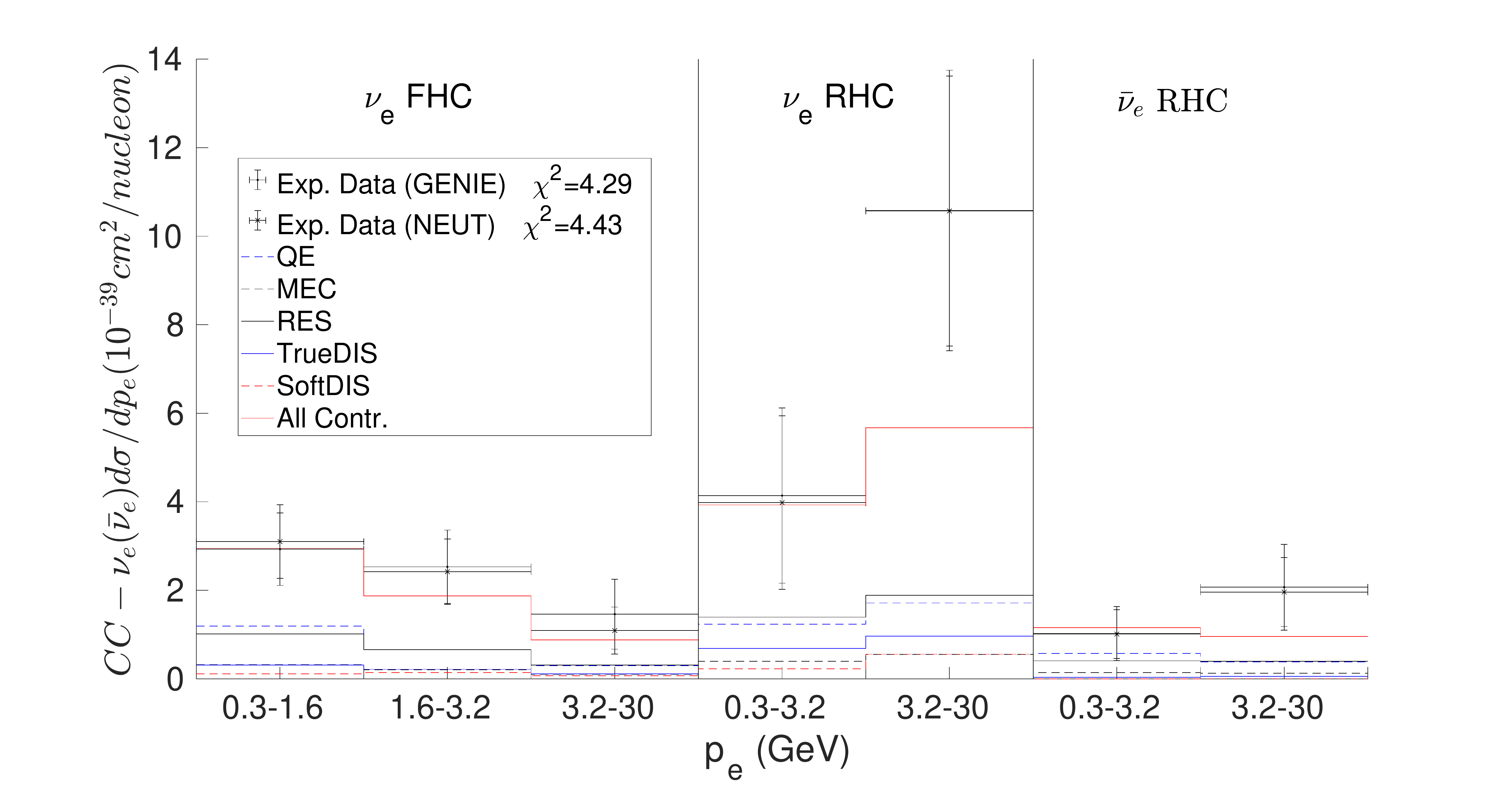}
    \caption{T2K CC electron neutrino and antineutrino inclusive flux-averaged total cross section per target nucleon  as function of the electron momentum. Different fluxes are used: forward horn current (FHC) and reversal horn current (RHC). Legend as in previous figures 
    (see Table~\ref{Table}). Data taken from~\cite{abe_measurement_2020-1}.
    \label{T2K_totalcs} }
\end{figure*}

In Fig.~\ref{T2K_totalcs} we compare our prediction for the total electron neutrino and antineutrino cross section on carbon against the electron momentum $p_{e}$ with data corresponding to forward (FHC) and reversal (RHC) horn current flux. The FHC electron  neutrino flux peaks at $\sim$ 1.2 GeV, whereas the RHC (anti)neutrino flux peaks at $\sim$(0.85)1.95 GeV. As observed, the data below $p_{e}=3$ GeV match our predictions whereas at higher electron momentum our models clearly underestimate data for most kinematics. The values of $\chi^{2}$ corresponding to NEUT and GENIE data are very similar and slightly smaller than the ones given in \cite{abe_measurement_2020-1}.


\subsubsection{MINERvA}

In this section we analyze the double and single differential flux-folded $\nu_{\mu}$-$^{12}$C inclusive cross sections per target nucleon versus the longitudinal ($p_{L}\equiv p_{\mu}\cos\theta_{\mu}$) and transverse ($p_{T} \equiv p_{\mu}\sin\theta_{\mu}$) momentum of the muon.

The target for the MINERvA experiment is hydrocarbon and the angular acceptance is limited to $\theta_{\mu}<20^{\circ}$. There are two sets of data corresponding to the two MINERvA fluxes, called low energy (LE) and medium energy (ME) fluxes. In the LE case the neutrino energy flux is peaked at $\sim 3.5$ GeV and the muon momentum is limited to $1.5<p_{L}/{\rm GeV}<20$, $p_{T}<2.5$ GeV. In the ME case the neutrino flux peaks at $\sim 6$ GeV and the muon momentum is limited to $1.5<p_{L}/{\rm GeV}<60$, $p_{T}<4.5$ GeV.    

\begin{figure*}[!htbp]
  \includegraphics[width=\textwidth]{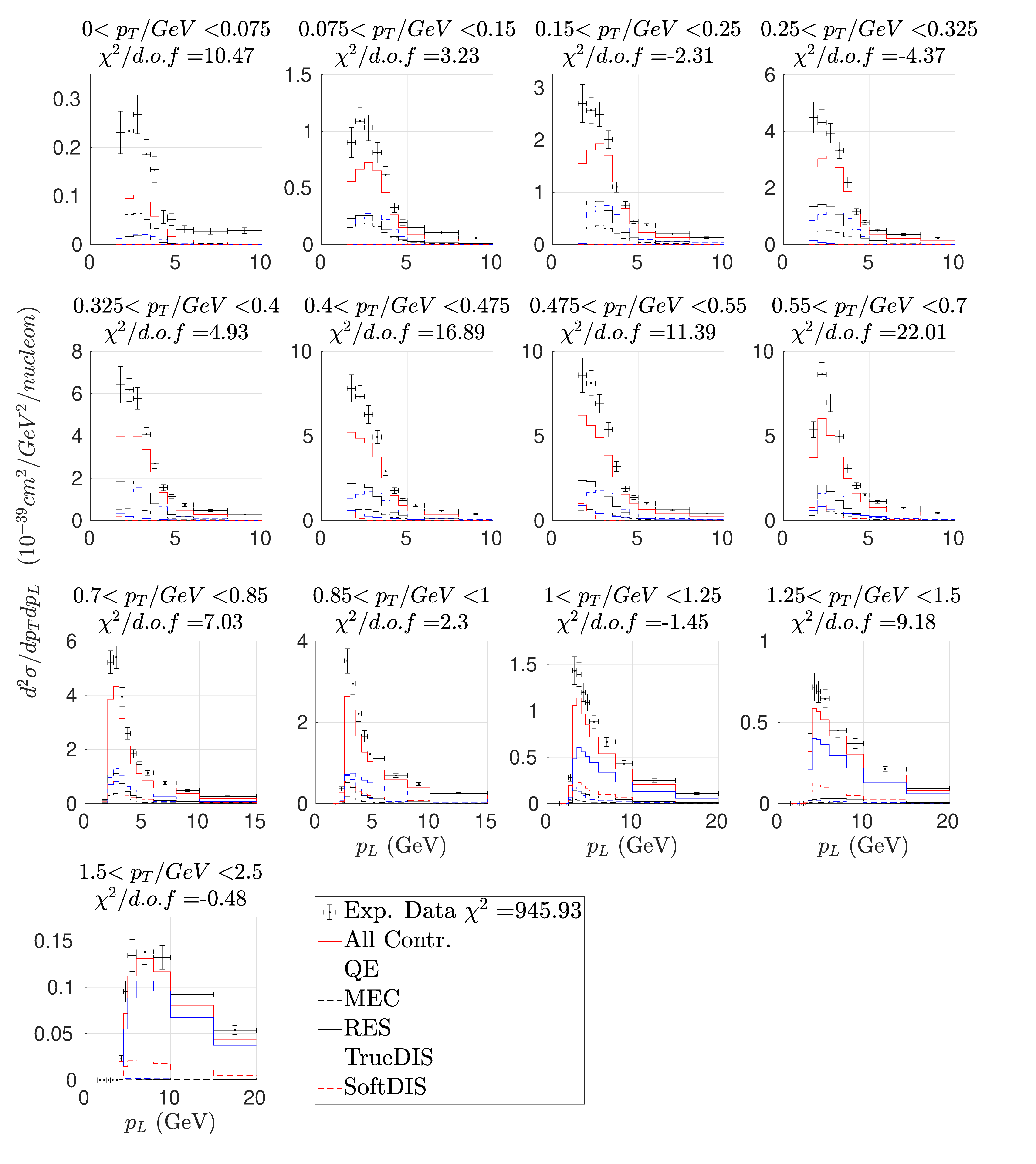}
    \caption{MINERvA CC inclusive (LE) flux-averaged double-differential cross section per target nucleon in bins of the transverse momentum as function of the longitudinal momentum. Legend as in previous figures (see Table \ref{Table}). Data taken from \cite{minera_collaboration_double-differential_2020}. The $\chi^{2}$- value shown in each panel is a partial calculation associated to each bin. We are using Eq. \ref{eq:chi2} to calculate the result portrait in legend.
    \label{Minerva_dpl_low} }
\end{figure*}

\begin{figure*}[!htbp]
  \includegraphics[width=\textwidth]{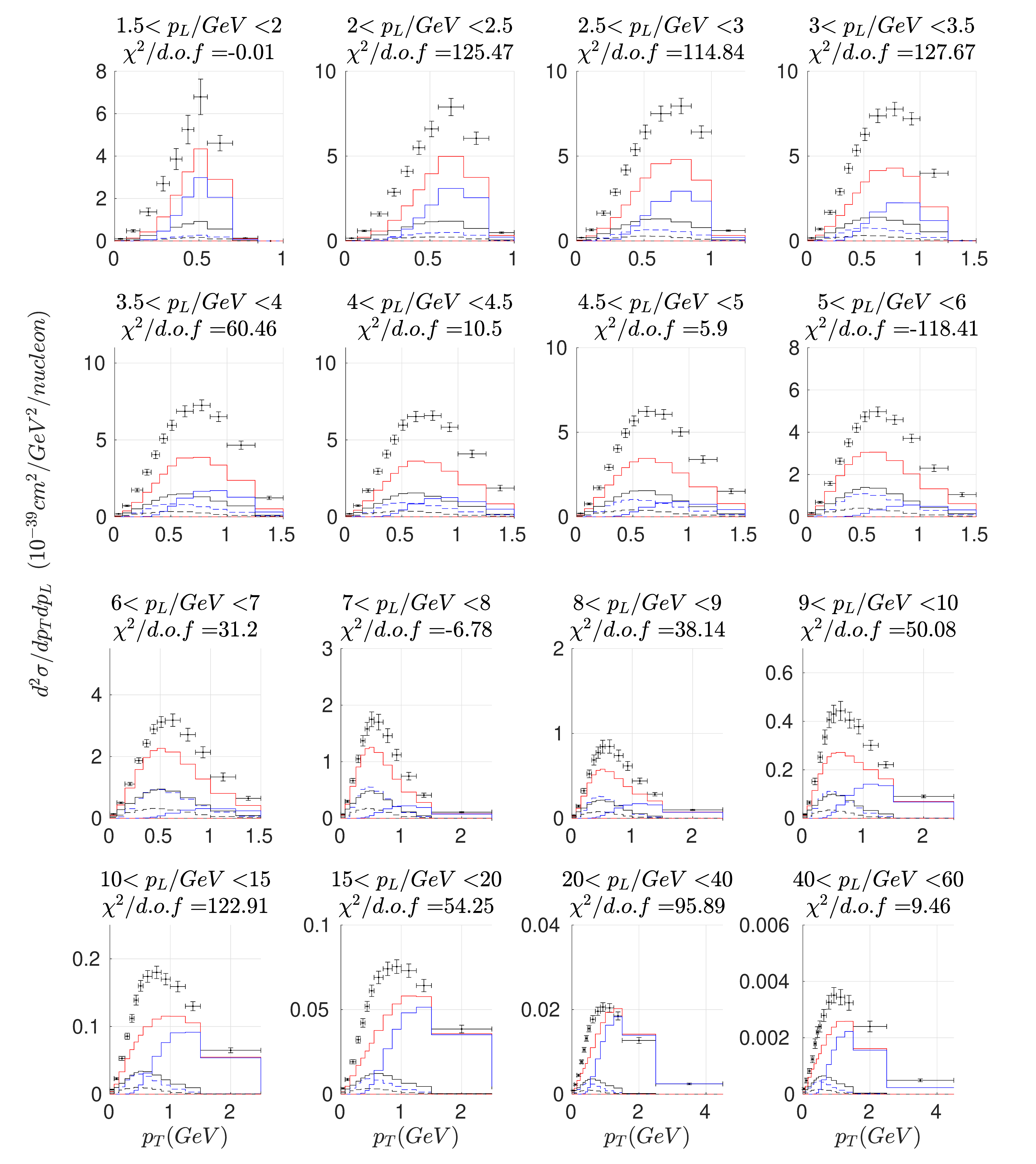}
    \caption{MINERvA CC inclusive (LE) flux-averaged double-differential cross section per target nucleon in bins of the longitudinal momentum as function of the transverse momentum. Legend as in previous figures (see Table~\ref{Table}). Same legend as Fig.~\ref{Minerva_dpl_low}. Data taken from \cite{minera_collaboration_double-differential_2020}.  The $\chi^{2}$- value shown in each panel is a partial calculation associated to each bin.
    \label{Minerva_dpt_low} }
\end{figure*}

In Figs.~\ref{Minerva_dpl_low} and \ref{Minerva_dpt_low}, the double differential cross section folded by the LE flux is presented as a function of $p_L$ and $p_T$, respectively. At low $p_T$, QE scattering gives the largest contribution to the cross section, followed by RES and MEC. As the transverse momentum increases, TrueDIS becomes more and more important, overcoming  the RES contribution. Compared to MINERvA GENIE predictions \cite{minera_collaboration_double-differential_2020}, we lack strength in the RES contribution, but get similar results for the other terms. This disagreement in the RES predictions might be connected with the way in which nuclear effects have been implemented in the SuSAv2 inelastic scaling function. Another source of differences can be linked to the Rein-Sehgal pion production model used in MonteCarlo simulations. This is an oversimplified description of the resonance contribution that cannot reproduce accurately the behavior of the cross section. This topic is discussed again later.
On the other hand, our $\chi^2$-analysis gets higher values than the ones presented by the simulations. In the figures we show the $\chi^2$-values divided by the number of data points, so they can be compared directly for the different kinematical situations analyzed.

\begin{figure*}[!htbp]
  \includegraphics[width=\textwidth]{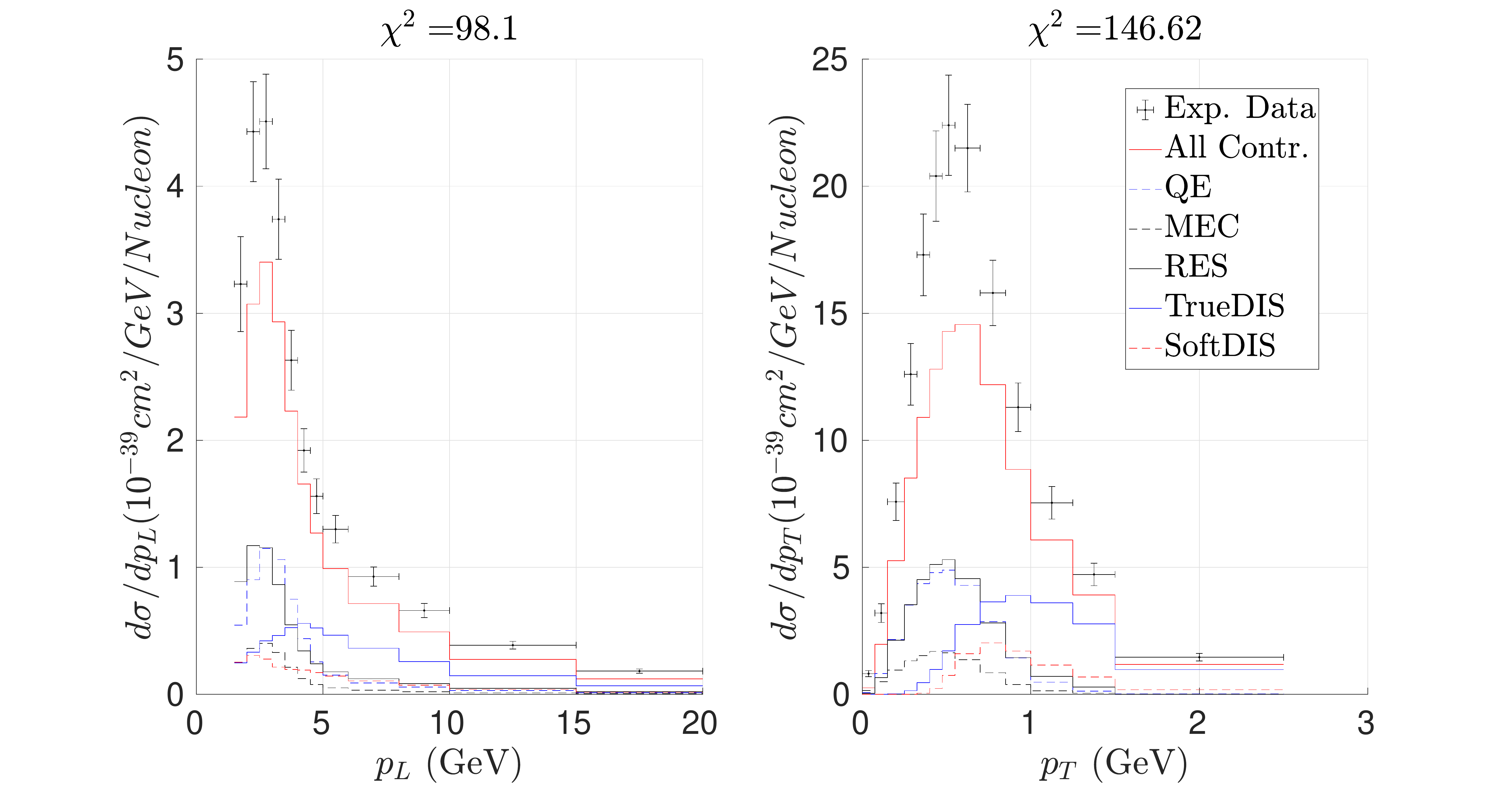}
    \caption{MINERvA CC inclusive (LE) flux-averaged single-differential cross section per target nucleon as function of the longitudinal momentum (left) and the transverse momentum (right) . Legend as in previous figures (see Table \ref{Table}). Data taken from \cite{minera_collaboration_double-differential_2020}.  The $\chi^{2}$-values are calculated using Eq. \ref{chi2-sc}.
    \label{Minerva_sc_low} }
\end{figure*}

In Fig.~\ref{Minerva_sc_low}, we show the LE single-differential cross sections versus $p_L$ (left panel) and $p_T$ (right). According to our prediction, QE and MEC account for around $30\%$ of the total strength. This is similar to the contribution ascribed to the TrueDIS channel, while SoftDIS only provides about $10 \%$ of the total cross section. Notice that, as in the previous case, we underestimate the data.  


\begin{figure*}[!htbp]
  \includegraphics[width=0.90\textwidth]{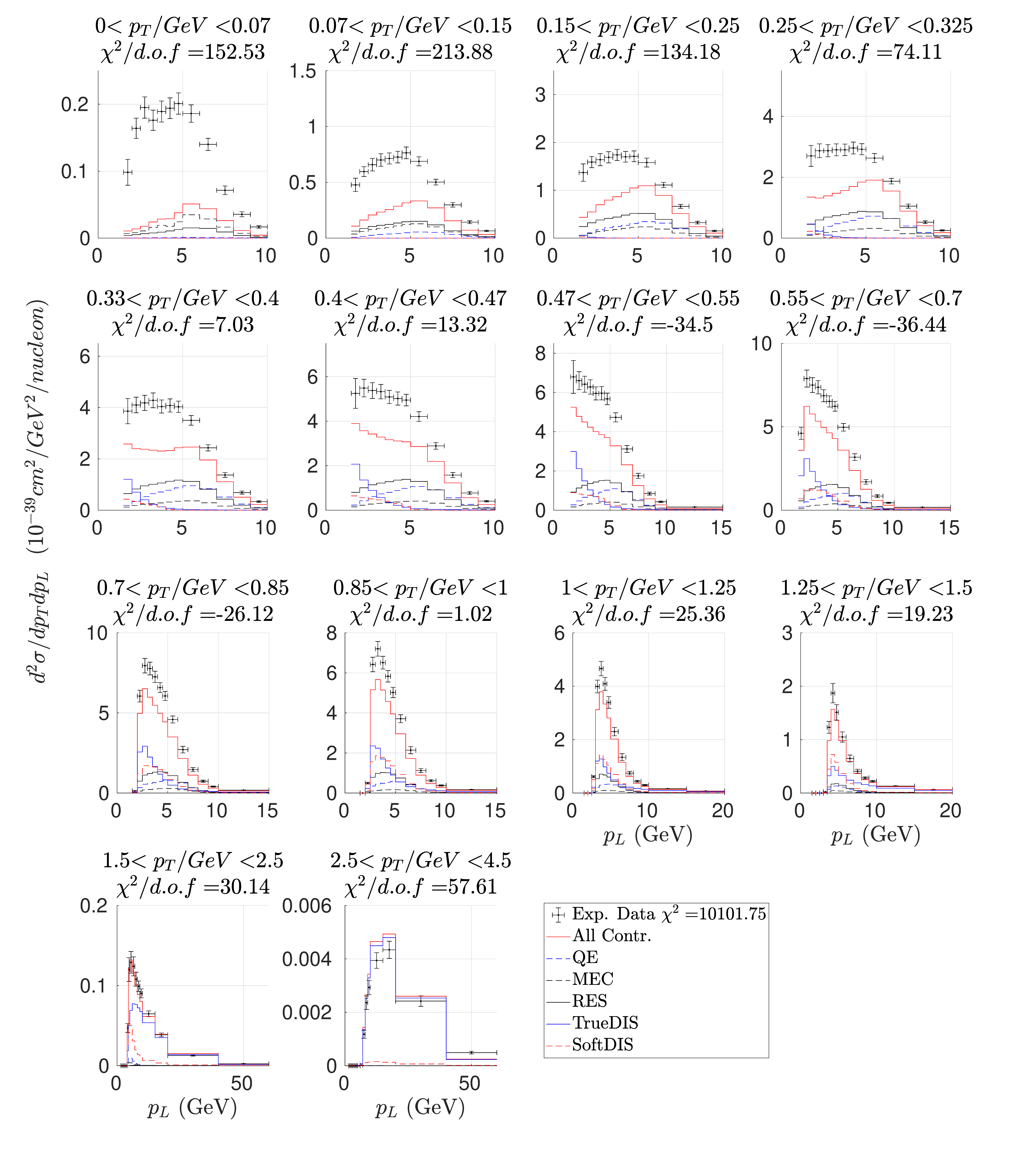}
    \caption{MINERvA CC inclusive (ME) flux-averaged double-differential cross section per target nucleon in bins of the transverse momentum as function of the longitudinal momentum. Legend as in previous figures (see Table \ref{Table}). Data taken from \cite{ruterbories_measurement_2021}. The $\chi^{2}$- value shown in each panel is a partial calculation associated to each bin. We are using Eq. \ref{eq:chi2} to calculate the result portrait in legend. \label{Minerva_dpl_medium} }
\end{figure*}

\begin{figure*}[!htbp]
  \includegraphics[width=\textwidth]{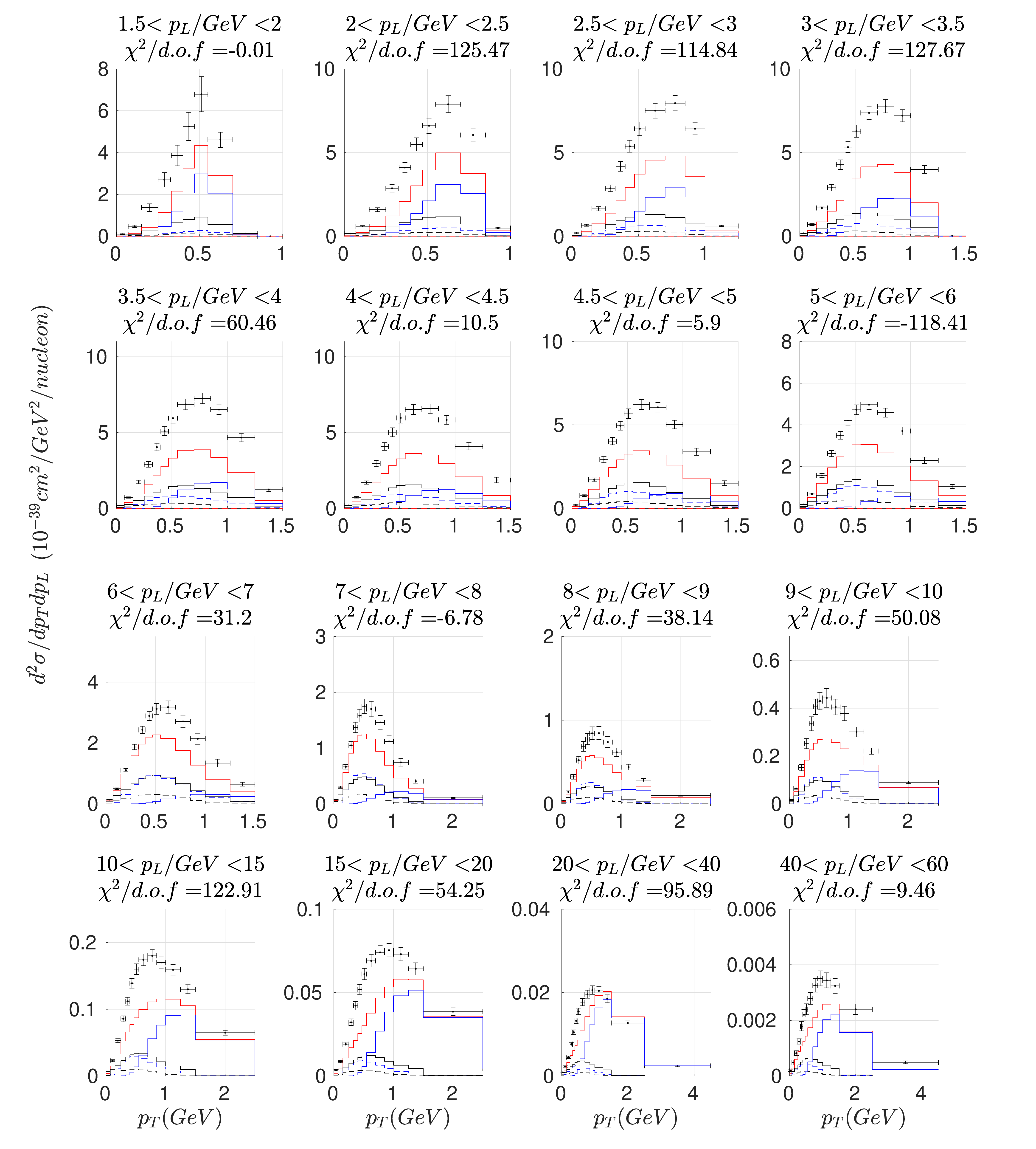}
     \caption{MINERvA CC inclusive (ME) flux-averaged double-differential cross section per target nucleon in bins of the longitudinal momentum as function of the transverse momentum. Legend as in previous figures (see Table \ref{Table}). Same legend as Fig. \ref{Minerva_dpl_medium} taken from  \cite{ruterbories_measurement_2021}. The $\chi^{2}$- value shown in each panel is a partial calculation associated to each bin. \label{Minerva_dpt_medium} }
\end{figure*}

 In Figs.~\ref{Minerva_dpl_medium} and \ref{Minerva_dpt_medium}, we show the double differential cross section folded by ME flux. As observed, the results and their comparison with data are very similar to the ones for the LE flux. In this case, compared to the analysis presented in \cite{minera_collaboration_double-differential_2020}, we lack strength in the SoftDIS contribution, and we tend to underestimate data by $\sim 20\%$. 
 The $\chi^{2}$-values are larger than the ones presented by the models used in MINERvA simulations and the ones shown for the LE flux.

\begin{figure*}[!htbp]
  \includegraphics[width=\textwidth]{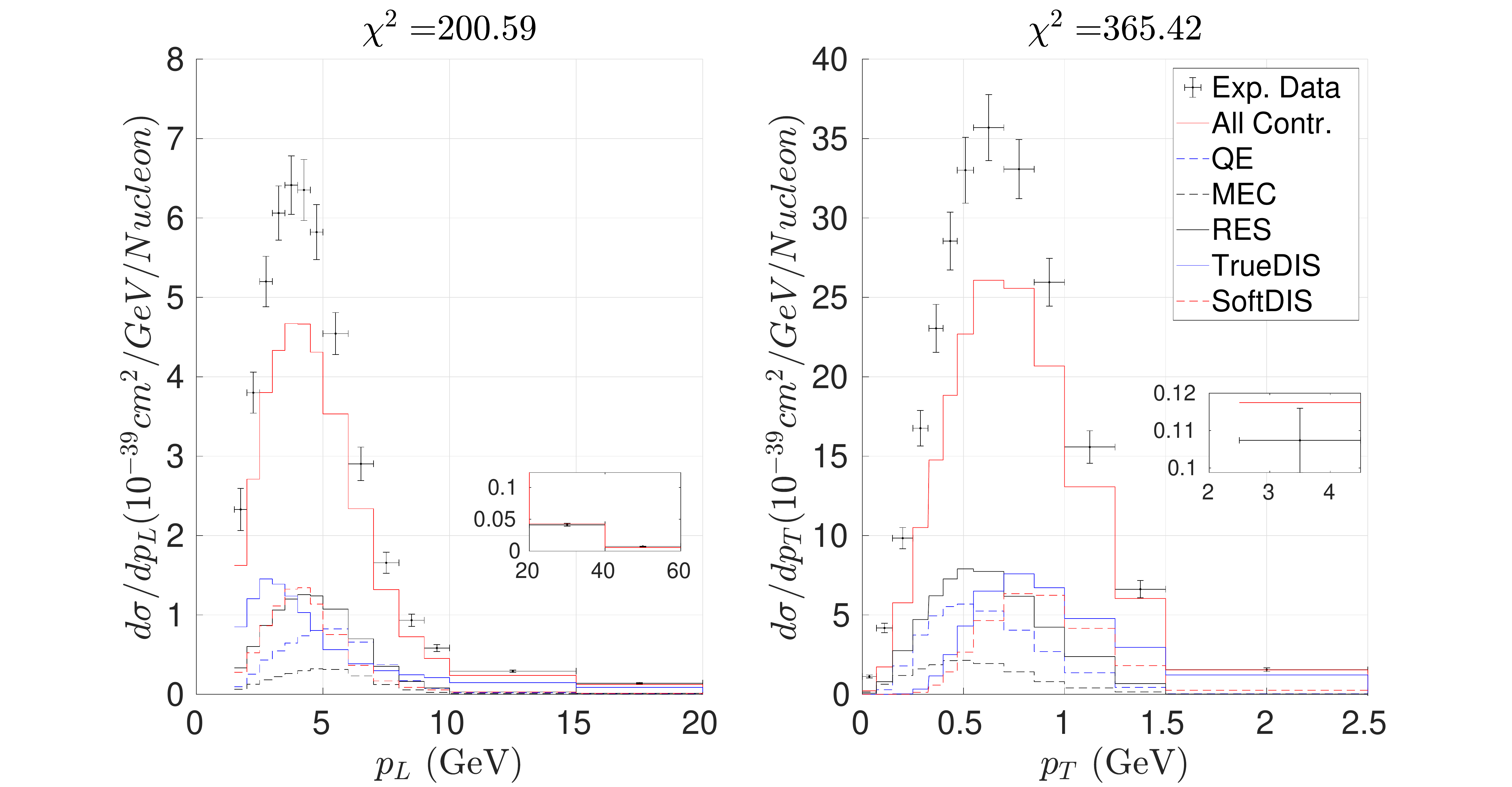}
    \caption{MINERvA CC inclusive (ME) flux-averaged single-differential cross section per target nucleon as function of the longitudinal momentum (left) and the transversal momentum (right). Legend as in previous figures (see Table \ref{Table}). Data taken from  \cite{ruterbories_measurement_2021}.  The $\chi^{2}$-values are calculated using Eq. \ref{chi2-sc}.
    \label{Minerva_sc_medium} }
\end{figure*}

Finally, in Fig.~\ref{Minerva_sc_medium} the single-differential cross section folded by the ME flux is shown. QE and MEC contributions are around $25 \%$, being the inelastic channels the ones that dominate at these kinematics. TrueDIS and SoftDIS contributions provide around half of the total strength of the cross section. As in previous cases, our predictions clearly underestimate data by a factor $\sim 1,4-1,3$ in the region of the maximum.

\begin{figure*}[!htbp]
  \includegraphics[width=\textwidth]{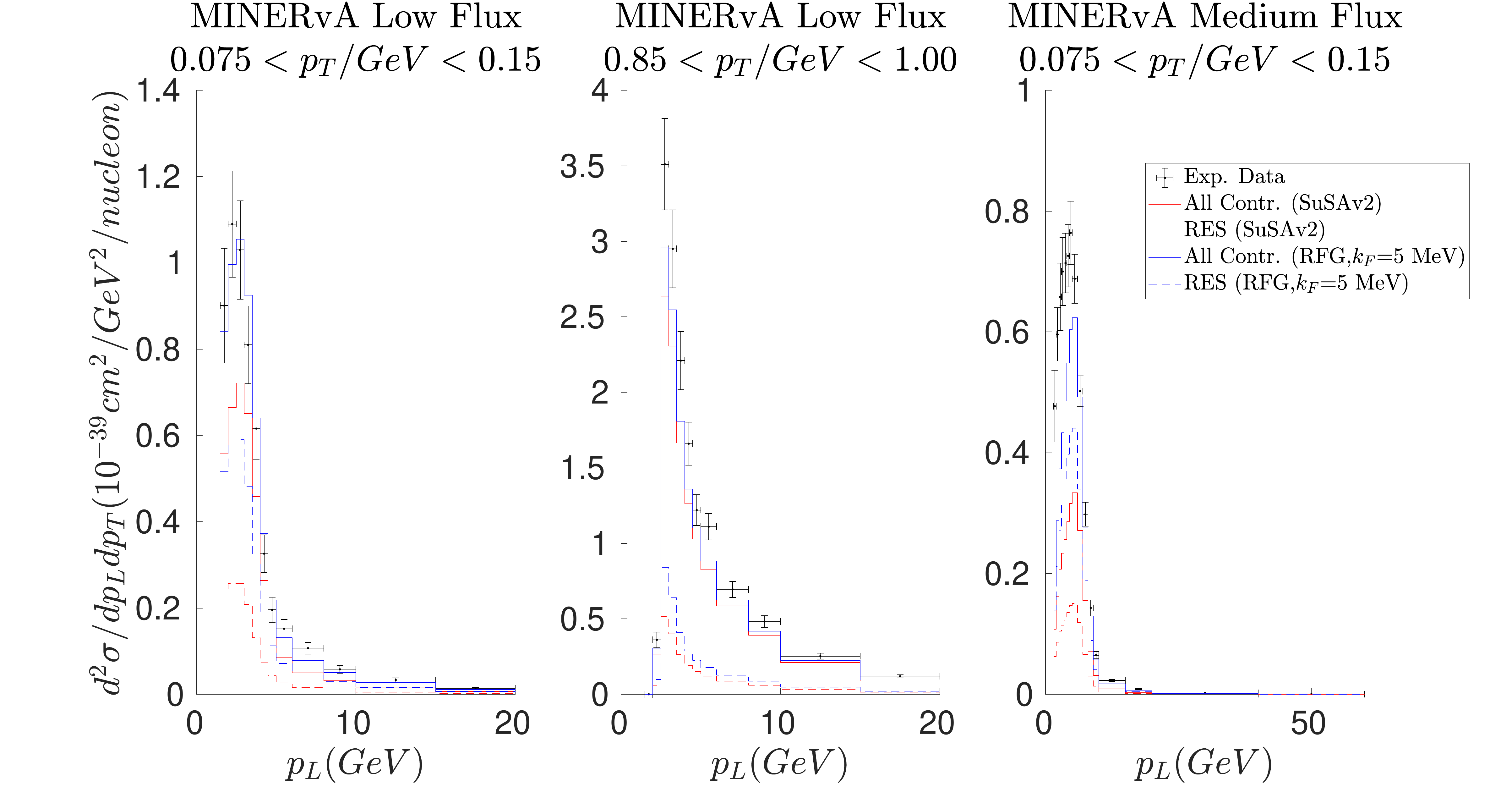}
    \caption{MINERvA CC inclusive (LE and ME) flux-averaged double-differential cross section per target nucleon as function of the longitudinal momentum using SuSAv2 and RFG (with $k_{F}=5$  MeV) scaling function. Legend as in previous figures (see Table \ref{Table}). Data taken from  \cite{ruterbories_measurement_2021}.
    \label{Minerva_sc_lowmedium} }
\end{figure*}

 This source of discrepancy between our predictions and data analysis based on GENIE simulations is better illustrated in the results presented in Fig.~\ref{Minerva_sc_lowmedium}. Here we explore the case of using very low Fermi momentum ($k_F=5$ MeV/c), and compare SuSAv2 inelastic and pure RFG predictions. Notice that the later agree much better with data. It is important to point out that the RFG predictions at very small $k_F$-values mimic the single-nucleon prediction, {\it i.e.,} nuclear effects are dismissed. Hence, the discrepancy between our two models is clearly connected with the nuclear effects introduced within the SuSAv2 approach. This is consistent with the resonance model implemented in MINERvA GENIE~\cite{GENIE:2021npt}, that is based on a single-nucleon Rein-Sehgal pion production model with lepton mass corrections. This approach increases the original Rein-Sehgal RES contribution by $\sim 20-40\%$, that seems to coincide with the discrepancies observed in 
Fig.~\ref{Minerva_sc_lowmedium}. However, further studies are needed to clarify this topic.



\subsubsection{MicroBooNE}

\begin{figure*}[!htbp]
  \includegraphics[width=\textwidth]{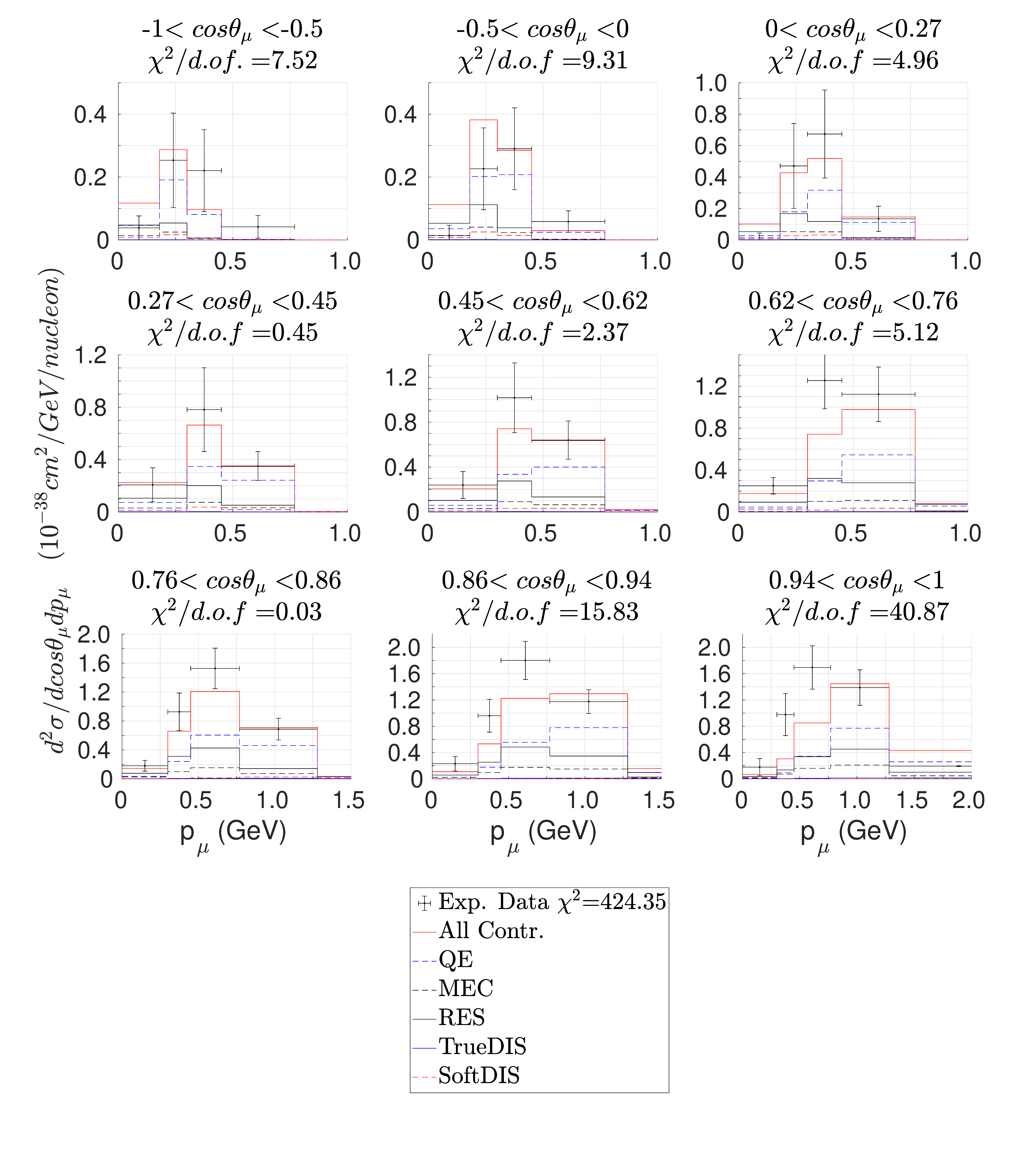}
    \caption{MicroBooNE CC inclusive flux-averaged double-differential cross section per target nucleon in bins of the muon scattering angle as function of the muon momentum. Legend as in previous figures (see Table \ref{Table}). Data taken from \cite{abratenko_first_2019-1}. The $\chi^{2}$- value shown in each panel is a partial calculation associated to each bin. We are using Eq. \ref{eq:chi2} to calculate the result portrait in legend.
    \label{Microboone} }
\end{figure*}

The MicroBooNE neutrino beam flux is peaked at $\sim$ 0.8 GeV and the target is liquid argon. In Fig.~\ref{Microboone}, we show the CC-inclusive $\nu_{\mu}$-$^{40}$Ar double differential cross section versus the muon momentum for different scattering angle bins. In each case we also present the $\chi^2$ analysis. As observed, the discussion of the results follows similar trends to the ones applied to T2K. On one hand, QE plus MEC contributions dominate at all kinematics, being around $70 \%$ of the cross section. On the other, the RES contribution provides around $20-30\%$ of the total strength, and it increases as the scattering angle gets larger. Finally, TrueDIS and SoftDIS contributions are not very relevant for these kinematics, being below $6\%$ of the cross section in all cases. 
Furthermore, the cross section shifts at lower momentum at very backward angles and the opposite is observed at very backward angles. The predictions explain rather well the data, except for the lowest scattering angles, for which the predicted shift is too strong.
Finally, we should note that the $\chi^{2}$-studies for the simulations are better than our analysis.

\subsubsection{ArgoNEUT}
\begin{figure*}[!htbp]
  \includegraphics[width=\textwidth]{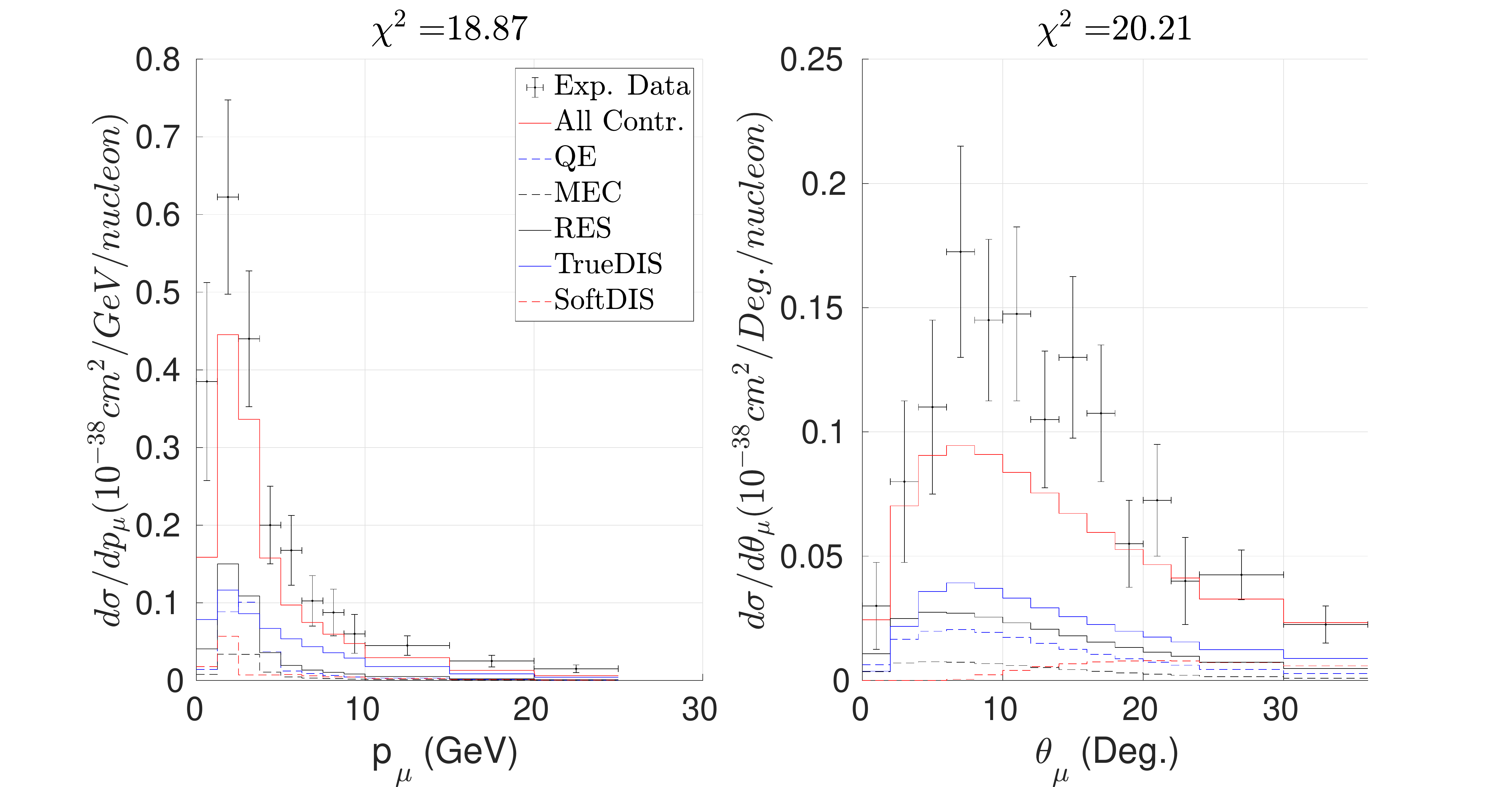}
    \caption{ArgoNEUT (2012) CC inclusive flux-averaged single-differential  neutrino cross section per target nucleon as function of the muon momentum (left) and as a function of the scattering angle (right). Legend as in previous figures (see Table \ref{Table}). Data taken from \cite{argoneut_collaboration_first_2012}.  The $\chi^{2}$-values are calculated using Eq. \ref{chi2-sc}.
    \label{ArgoNEUT_old} }
\end{figure*}

\begin{figure*}[!htbp]
  \includegraphics[width=\textwidth]{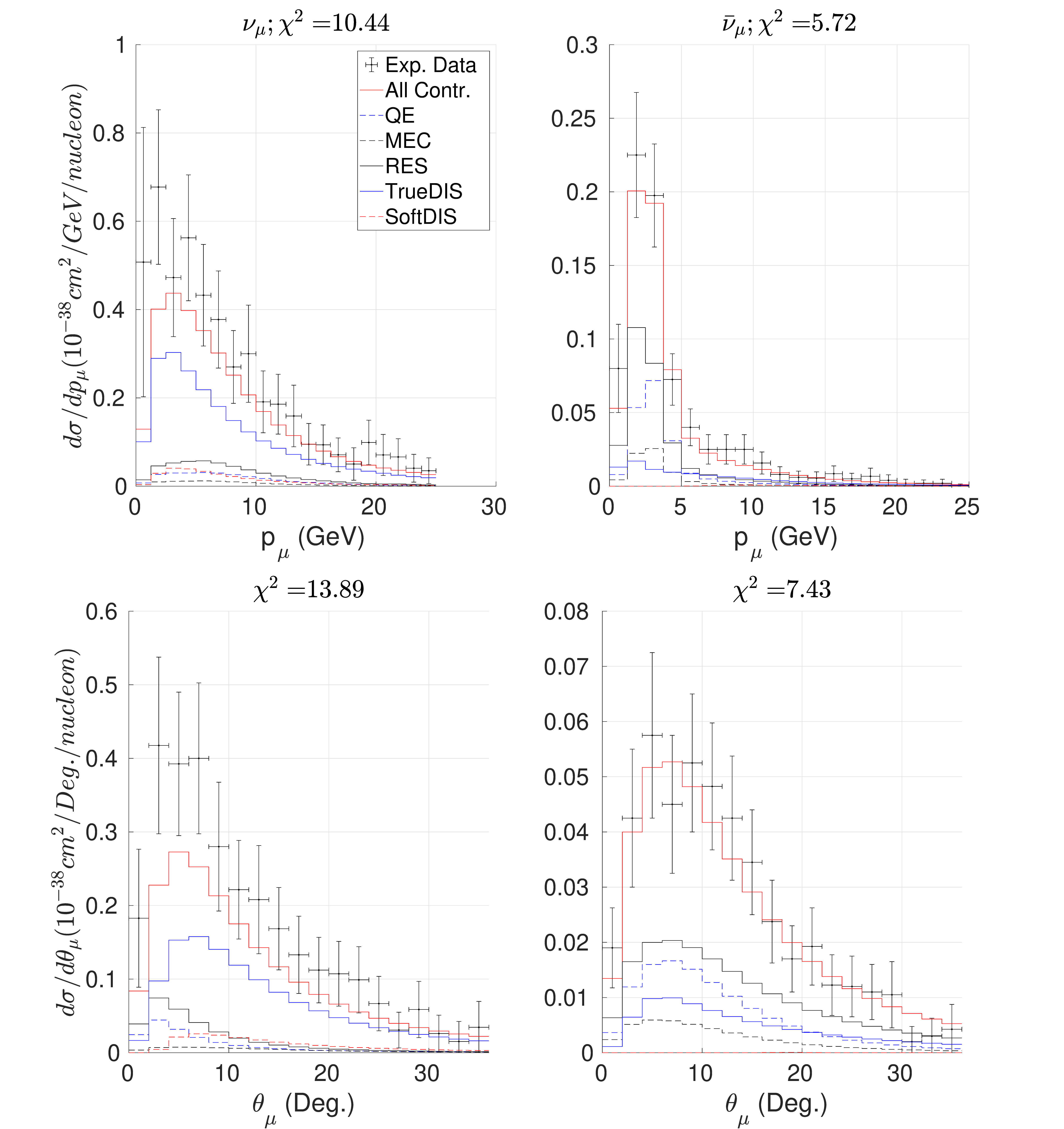}
    \caption{ArgoNEUT (2014) CC inclusive flux-averaged single-differential cross section per target nucleon as function of the muon momentum (top) for neutrinos (left) and antineutrinos (right) and as a function of the scattering angle (bottom). Legend as in previous figures (see Table \ref{Table}). Data taken from \cite{acciarri_measurements_2014-2}.  The $\chi^{2}$-values are calculated using Eq. \ref{chi2-sc}.
    \label{ArgoNEUT_new} }
\end{figure*}

In the ArgoNEUT experiment, as for MicroBooNE, the target is liquid argon. The ArgoNEUT(2012) neutrino flux peaks at 4.5 GeV. The (anti)neutrino ArgoNEUT(2014) flux is peaked at (3.6) 9.6 GeV. In Figs.~\ref{ArgoNEUT_old} and \ref{ArgoNEUT_new}, the CC-inclusive $\nu_{\mu}$-$^{40}$Ar flux-folded single differential cross section is shown as a function of the muon momentum $p_{\mu}$ (left panel in  Fig.~\ref{ArgoNEUT_old} and top panels in Fig.~\ref{ArgoNEUT_new}) and the scattering angle $\theta_{\mu}$ (right panel in Fig.~\ref{ArgoNEUT_old}, bottom panels in Fig.~\ref{ArgoNEUT_new}). The acceptance is $\theta_{\mu}<36^{\circ}$ and the muon momentum is limited to $p_{\mu}<25$ GeV.
For neutrinos (Fig.~\ref{ArgoNEUT_old} and left panels in Fig.~\ref{ArgoNEUT_new}), it is observed that QE, MEC and RES provide $\sim 50\%$  of the strength of the cross section for ArgoNEUT (2012). Meanwhile, for ArgoNEUT (2014), TrueDIS being the dominant contribution by giving $\sim 65 \%$  of the strength of the cross section. Notice that our predictions underestimate data, particularly in the region where the cross section reaches its maximum. Nevertheless, the models are capable of explaining the fall of the cross section. 

In the case of antineutrinos (right panels in Fig.~\ref{ArgoNEUT_new}), the models reproduce the cross section data perfectly well. It is worth noticing that the flux is different for antineutrinos and neutrinos, peaking at different energies. In the case of antineutrinos, QE plus MEC provides $\sim 40\%$ of the total strength, whereas the inelastic channels are dominated by the RES contribution. This is clearly in contrast to the results observed for neutrinos, and it is linked to the nice accordance between our theoretical predictions and data for antineutrinos, whereas they depart significantly in the case of neutrinos. However, further studies are needed to explain which basic ingredients in the analysis of ArgoNEUT (anti)neutrino-nucleus scattering are responsible of the differences observed in both cases.

\section{Conclusions} \label{Conclusions}

In this work we have expanded upon the SuSAv2-inelastic model, presented in previous work~\cite{gonzalez-rosa_susav2_2022}, by  implementing the Dynamical Coupled Channels model~\cite{nakamura_dynamical_2015-3}, which takes into account all the possible resonances of the nucleon and their mutual interactions involving  pions, double pions, kaons, etc. This represents a clear improvement of the previous model, in which the resonance region was described by phenomenological fits. However
the limitation of the DCC model to resonance production makes it necessary to add the deep inelastic scattering contribution, which may occur in the same kinematic region: this we do by using the original SUSAv2-inelastic model with appropriate kinematic cuts. The two models complement each other to obtain what we define as SoftDIS contribution. The rest of the inelastic spectrum is described by the SuSAv2-inelastic model alone. 

This new iteration of the model has been tested against electron-carbon scattering data with excellent results in a wide range of kinematics. After that, the model has been applied to neutrino scattering, and compared its predictions with several CC-inclusive data. 

For T2K and MicroBooNE the model is capable to explain the data really well. The discrepancies between predictions and data from these experiments are very similar to those found in our previous work~\cite{gonzalez-rosa_susav2_2022} and, particularly in the case of T2K, they are not related to the resonance contribution. 

In the case of MINERvA and ArgoNEUT, which operate at higher energies, our results tend to underestimate the experimental data. In general, for the low energy MINERvA flux, the resonance contribution lacks the strength shown by Monte Carlo simulations~\cite{minera_collaboration_double-differential_2020}, while SoftDIS lacks strength for the medium energy MINERvA flux. In the results for neutrinos from ArgoNEUT, we also lack strength in general in comparison with data. However, the antineutrino ArgoNEUT data, corresponding to lower energy, are perfectly explained by the model. 

In comparing with experimental data we have performed a $\chi^{2}$ analysis in order to quantify the quality of the agreement and compare it with the one reported in the experimental papers and obtained by Monte Carlo simulations. We have found that, in general, the SuSAv2 model gives higher values of $\chi^{2}$  than those obtained by the simulations. 
These discrepancies need further investigation. However, it should be stressed that our model does not contain any free parameter: the few parameters have been fixed once and for all by comparing with electron scattering data~\cite{megias_inclusive_2016-1}. Therefore, no "tuning" of the model is performed to adjust to neutrino data.

In future works, we plan to implement other resonance models~\cite{Kabirnezhad_2018} in the SuSAv2 framework. Moreover, the  recent inclusion of the DCC model in the NEUT simulator will allow for a direct comparison with our predictions.

\begin{acknowledgments}
This work was partially supported by the Spanish Ministerio de Ciencia, Innovación y Universidades and ERDF (European Regional Development Fund) under contract PID2020-114687GB-100,  by the Junta de Andalucia (grants No.~FQM160, SOMM17/6105/UGR and P20-01247), by University of Tokyo ICRR’s Inter-University Research Program FY2022 (Ref.~2022i-J-001) $\&$ FY2023 (Ref.~2023i-J-001),  by the INFN under project Iniziativa Specifica NucSys and the University of Turin under Project BARM-RILO-20 (M.B.B.). J.G.R. was supported by a Contract PIF VI-PPITUS 2020 from the University of Seville (Plan Propio de Investigación y Transferencia) associated with the project PID2020-114687GB-100.
\end{acknowledgments}

\bibliography{DCC.bib}

\end{document}